# Similarity Search Combining Query Relaxation and Diversification

Ruoxi Shi, Hongzhi Wang, Tao Wang, Yutai Hou,

Yiwen Tang, Jianzhong Li, Hong Gao

Harbin Institute of Technology, Harbin, China
`{shiruoxi,wangzh,lijzh,honggao}`@hit.edu.cn yt6789299@163.com
`{atma.hou,isabeltang147}`@gmail.com

**Abstract.** We study the similarity search problem which aims to find the similar query results according to a set of given data and a query string. To balance the result number and result quality, we combine query result diversity with query relaxation. Relaxation guarantees the number of the query results, returning more relevant elements to the query if the results are too few, while the diversity tries to reduce the similarity among the returned results. By making a trade-off of similarity and diversity, we improve the user experience. To achieve this goal, we define a novel goal function combining similarity and diversity. Aiming at this goal, we propose three algorithms. Among them, algorithms genGreedy and genCluster perform relaxation first and select part of the candidates to diversify. The third algorithm CB2S splits the dataset into smaller pieces using the clustering algorithm of k-means and processes queries in several small sets to retrieve more diverse results. The balance of similarity and diversity is determined through setting a threshold, which has a default value and can be adjusted according to users' preference. The performance and efficiency of our system are demonstrated through extensive experiments based on various datasets.

## 1    Introduction

Similarity search that finds objects with distance larger than a given similarity threshold or within a certain distance threshold with the query in a dataset has a wide range of applications, such as web page detection, entity linking and protein identification[13][14][15].

Recently, the quality of similarity search results has attracted more attention. The result quality is often measured in two dimensions.

One is the number of results. Too few results provide insufficient results to the user, while too many results are inefficient to display and impossible for users to explore. When too few results are returned, the query has to be relaxed to obtain more results. For example, wrong or fuzzy input may cause few searching results if the keyword is "Briatney". The searching engine will obtain more results by correcting the keyword to "Britney", which is the name of a famous singer.

The other is the diversification of results, which is the quantitative description of the variety of elements in the result set. A good search engine attempts to provide various kinds of information within limited number of results. In web search engines and recommendation systems, query result diversification helps counteract the over-specialization problem in which the retrieved results are too homogeneous to meet users' needs [16][17][3].

During the similarity search process, these two dimensions are correlative and should be balanced. We use an example to illustrate this point. Consider the scenario of searching for commodities in an e-commerce site. The best search result is to show users abundant but not redundant commodities. These commodities meet the requirement of user input and meanwhile, different enough to one another. Similarly, such technique can also be applied to information retrieval, image search and some other areas [28] [29].

Such requirements bring challenges to query processing to obtain high-quality results. To find the most diverse elements has been proved to be an NP-complete problem [20], and one optimal solution leads to incredible time and space cost, especially on massive and complex datasets.

Even though many query relaxation and result diversification approaches have been proposed, they fail to balance result number and diversification. Relaxation techniques in [21] [22] only perform relaxation when the query result is empty, and the result number is uncontrollable. Moreover, the similarity among the results gets high due to the relaxation. Algorithms in [3] [4] [5] return $k$ diverse neighbors, and most of them have a two-step of candidate-filter selection based on greedy selection. Due to the facts that the optimality of greedy selection is not guaranteed, and the result quality of candidate-filter algorithms is greatly affected by the quality of the candidate set, a bad candidate set may lead to worse results.

In this paper, we attempt to obtain a proper number of results with high diversity. We control the result number within a range $[k_{min}, k_{max}]$ instead of a fixed integer $k$ in previous studies [7] [19]. In practice, the lower bound is often given by the user, while the upper bound is limited to the result display interface or the user's ability of exploring the result. With the consideration of relaxation, we define the problem with the measure combining diversity and similarity and develop various algorithms based on this measure.

For different scenarios, we develop three algorithms to solve this problem. genGreedy is based on greedy selection strategy, with high efficiency, and is more applicable for a frequently changing dataset. Based on multiple sequence alignment, genCluster costs more time than genGreedy, but more stable. The third algorithm CB2S is based on cluster analysis and machine learning. It is designed to achieve high efficiency aiming at complex and massive datasets.

The contributions of this paper are as follows.

— We study efficient query processing with the consideration of both result number and diversification. As far as we know, this is the first paper considering both of these dimensions.
— To achieve the goal, we design a novel measure of query result quality combining similarity and diversity. We develop three efficient algorithms for different scenarios.
— For efficient query processing, we develop a string vectorization strategy and iterative query processing strategy to speed up the search process.
— We tested our approaches on various real datasets. Extensive experimental results show that when returning similar diversity with existing algorithms, our approach provides a proper number of results. The runtime comparison shows that our approaches are more efficient.

The rest of the paper is organized as follows. Problem definition is discussed in Section 2. Section 3, 4 and 5 describes our three searching approaches in detail. Our experimental results are presented in Section 6, and we conclude our paper in Section 7.

## 2  Problem definition

In this section, we define the problem by defining the quality of query results integrating similarity between query and result, the number of results and diversification. For simplification, we focus on string and use edit distance [23] as the similarity measure. Our approaches can also be adapted to other applications such as semantic or image similarity search with minor changes on the search criteria.

We denote the dataset as $DS = \{s_1, s_2, ......, s_n\}$, the query as $q$, the given threshold of distance as $\varepsilon$ and the given result number range $[k_{min}, k_{max}]$.

**Definition 1.** Given a query $q$ and a result set $S$, the similarity between $q$ an $S$ is defined as the average distance between $q$ and all elements in $S$, i.e.
$$argSim(S,q) = \frac{1}{|S|} \sum_{s_i \in S} Dis(q, s_i)$$
where $Dis(q, s_i)$ is the distance between $q$ and $s_i$.

In this paper, we adopt content-based diversity based on edit distance, since the other two kinds (intent based diversity and novelty based diversity) are mainly used for semantic analysis [2]. The definition is as follows.

**Definition 2.** Given a result set $S$, the diversification of $S$ is defined as follows.
$$argDiv(S,q) = \frac{2}{k(k-1)} \sum_{s_i, s_j \in S} Dis(s_i, s_j)$$

Intuitively, the goal of query processing is to minimize the similarity distance and maximize the diversity. However, these two dimensions are correlative. To balance these two dimensions, we use a coefficient λ and define the objective function of the query process as follows.

**Definition 3.** Given a trade-off parameter of similarity and diversity, coefficient λ ( λ∈[0,1] ), the objective function $F(S,q)$ for a result set $S$ is as follows.
$$F(S,q) = \lambda\, argDiv(S,q) + (1-\lambda)\,(-argSim(S,q))$$

In this definition, the trade-off parameter λ can be determined by users. Hence the inner structure of returned set is flexible, i.e. a small λ leads to more relevant results while a large λ leads to more diverse results. Also, this parameter can be also decided by analyzing various datasets through model building or sampling like [20].

We chose the form $F(S,q)$ in three reasons. First, $F(S,q)$ is an efficient and effective assessment since it combines similarity and diversity into one expression and uses an adjustable parameter to balance these two dimensions. Furthermore, $F(S,q)$ increases with the growth of $argDiv(S,q)$ and drops with the growth of $argSim(S,q)$, which excellently reflects our aim at finding the most diverse results which are also similar to query $q$. Finally, this formula is simple and the computation cost is small.

According to this definition, the query processing algorithm works for a given query $q$ and range $[k_{min}, k_{max}]$, to retrieve a result set $S$ with $|S|=k \in [k_{min}, k_{max}]$ and maximize $F(S,q)$. According to [20], even when λ=0, this problem is NP-Complete. Thus, we attempt to design efficient heuristic algorithms in the following sections.

## 3  genGreedy

Intuitively, the proposed problem can be solved by two steps, generating sufficient candidates through relaxation and greedy selection. Based on this framework, we develop the query processing algorithm genGreedy.

This algorithm has two phases, candidate generation and diversification filter.

### 3.1  Candidate Generation

Candidate generation phase first generates $k$ results with the highest relevance with the query , which will be used for further selection. If the result number of an accurate query is smaller than $k$, relaxation is performed.

To ensure the number of final results, $k$ should be large enough, while to achieve high efficiency in the diversification filter phase and select the results similar enough to $q$, $k$ should not be too large. Hence in the relaxing process we make $k$ dependent on $k_{min}$ and $k_{max}$, $k \in [(\lambda+1)k_{min}, (\lambda+1)k_{max}]$, λ∈[0,1], as mentioned in section 2. Thus, $[k, 2k]$ strings are retrieved in the phase of relaxation, and in selecting phase, we pick $1/(\lambda+1)$ of the candidates since we enlarge the number constraint by $\lambda+1$ in relaxation.

To obtain the results which are the most similar to $q$, we develop an iterative algorithm for candidate generation phase. In this algorithm, the query is relaxed iteratively from the one most similar string to $q$ to those different ones until total $k$ results are obtained with the relaxed queries.

That is, if insufficient results are obtained through $q$ in the first round, then a greater threshold is used for query relaxation to retrieve results within difference $\varepsilon$ with $q$. Initially, $\varepsilon$ is set to 1 to retrieve the results within distance smaller than 1 with $q$. If such relaxation does not return sufficient results, $\varepsilon$ is relaxed to a larger value.

The pseudo code is shown in Algorithm 1. In this algorithm, for efficiency issues q-gram and inverted index [24] are adopted. We first initialize $min\_com$ with 0 and the output set $rlxResult$ with $\emptyset$ (Line 1-2). Line 3-12 describe the iterative process. In Line 3, we start the iteration until the result number equals or exceeds $(\lambda + 1) * k_{min}$. During each iteration, we turn to next string if the current string $s$ is already in $rlxResult$(Line 5-6). In each iteration, we set the value of $min\_com$, which is the minimum number of same grams that two strings should contain. Considering that the similarity of each result cannot be guaranteed to be within $\varepsilon$ if only one step of q-gram approach is used. Hence, we add a verification step in Line 8-9. The results that pass verification are added into the result set in Line 9. In Line 10, we check the number of results. The program jumps out the loop and returns set $rlxResult$ (Line 13) if $|rlxResult| = (\lambda + 1) * k_{max}$ is satisfied. If there are insufficient results, we enlarge $\varepsilon$ (Line 12) to perform the next round of searching.

Note that the computation of set similarity for the q-gram set of each string is inefficient, we involve inverted list to accelerate the process. The details will be discussed in Section 6.

**Algorithm 1: Relaxation**

**Input:**
  $q$: query string; $k_{min}, k_{max}$: minimum and maximum bound
  λ: tradeoff parameter; ε: threshold of edit distance

**Output:**
  $rlxResult$: set selected from dataset

1. $min\_com \leftarrow 0$
2. $|rlxResult| \leftarrow \emptyset$
3. **while** ($|rlxResult| < (\lambda + 1) * k_{min}$) **do**
4.   **for** (each string $s$ in $dataset$) **do**
5.     **if** $s$ in $rlxResult$
6.       **continue**
7.     $min\_com \leftarrow q.length + 2 - 1 - \varepsilon * 2$
8.     **if** $|s.\text{grams} \cap q.\text{grams}| > min\_com$ && $\text{Dis}(q,s) < \varepsilon$
9.       $rlxResult.add(s)$
10.    **if** $|rlxResult| = (\lambda + 1) * k_{max}$
11.      **break**
12.  $\varepsilon \leftarrow \varepsilon + 1$
13. **return** $rlxResult$

The complexity of Algorithm 1 is $O(k_{min}N)$, where $k_{min}$ is the minimum bound of the result number, and $N$ is the size of dataset $DS$.

### 3.2 Diversification Filter

In the diversification filter phase, we select top $(1/\lambda+1)* |rlxResult|$ strings that make the greatest contribution to result diversity. Since the diversity of a set is measured through $argDiv$ in Section 2, we define the contribution that a string $t$ makes to the final result set $S$, denoted by $DD_t(S)$, as follows.

**Definition 4.** Given two strings $s_i$ and $s_j$ in dataset $S$, the edit distance between them is $Dis(s_i, s_j)$. The contribution is computed as the sum of each distance between $t$ and any other string $s$, which is denoted as $DD_t(S) = \sum_{s \in S}^{n} Dis(s, t)$.

We accelerate the selection by pruning the strings that are not diverse enough ac-cording to a prune function $F(\sigma, \Omega) = \Omega \frac{1}{\sigma*|rlxResult|} \sum_{s \in rlxResult, t \in samSet}^{|rlxResult|} Dis(s,t)$ where $\sigma$ and $\Omega$ are the parameters of pruning, $\sigma \in (0,0.5)$ with default value of 0.25, $\Omega \in (0.5,1)$ with default value of 0.75. The sample set of $rlxResult$ is $samSet$ with size of $\sigma * |rlxResult|$. $\sigma$ decides how many elements $samSet$ contains. In order to guarantee the number of query results, we use parameter $\Omega$ to control pruning number. And $\Omega$ can be adjusted by users according to the preferences. Higher $\sigma$ and $\Omega$ increase the accuracy but decrease the efficiency of algorithm, and vice versa.

**Algorithm 2: greedy**

**Input:**
  $ED\_matrix$: two-dimensional matrix string edit distances between each pair of strings
  $rlxResult$: strings generated by Algorithm Relaxation ; λ: tradeoff parameter

**Output**:
  $S$: final result set

1. **for** each candidate $c_i$ in $rlxResult$ **do**
2. $\quad\quad DD_{c_i}(rlxResult) \leftarrow \sum_{j=0, j \neq i}^{n} ED\_matrix_{ij}$
3. calculate $F(\sigma, \Omega)$
4. **for** each candidate $c_i$ in $rlxResult$ **do**
5. $\quad\quad$ **if** $DD_{c_i}(rlxResult) < F(\sigma, \Omega)$
6. $\quad\quad\quad\quad rlxResult.\text{remove}(c_i)$
7. $S \leftarrow \text{mergeSort}(rlxResult)$
8. **return** $S$

The pseudo code is shown in Algorithm 2. Initially, for each candidate $c_i$ in $rlxResult$, we calculate how much contribution $c_i$ makes for $rlxResult$ by $DD_{c_i}(rlxResult)$ in Line 1 and 2. In Line 3, we calculate $F(\sigma, \Omega)$ for pruning in Line 5. Candidates with $DD_{c_i}(rlxResult)$ lower than $F(\sigma, \Omega)$ are considered to have a too low diversity and removed from $rlxResult$(Line 6). After this pruning, we sort the candidates by $DD_{node}(rlxResult)$ and return top $(1/λ+1) * |rlxResult|$ results as $S$ (Line 7 and 8).

The time complexity of Algorithm 2 is $O(k \log k)$, in which $k = |rlxResult|$. Since the cost of merge sort is $O(k \log k)$ and that of one loop is $O(k)$, the total complexity of genGreedy is $O(kN + k \log k)$.

This algorithm is simple and efficient without heavy preprocessing cost. Thus, it is suitable for scenarios with frequently changing datasets. As shown in Section 6, this algorithm could generate a good result set efficiently in most conditions, especially when dealing with complex and massive datasets. However, this method is not stable when datasets are too small. To remedy the shortage, we present a more stable approach genCluster in the next section.

## 4 genCluster

genCluster is presented to solve the unstable problem of genGreedy. As a trade-off, it relatively costs more time than genGreedy. Hence, it is more applicable for scenarios when the result quality requirement is more important than query runtime restriction.

To make the algorithm more stable, we cluster the candidates in $rlxResult$ based on multiple sequence alignment. Such idea is inspired by the method of multiple sequence alignment in bioinformatics [25], which finds genetic relation among series of DNA or proteins. We apply such idea to find similarity connection among strings. First of all, we make pairwise alignment to create a distance matrix. Thereafter, a guide tree is built by applying clustering algorithms. Then a motif string is created by a method of scoring. Finally, strings far away from the motif in edit distance are picked out to maximize the diversity. This algorithm can also be divided into two parts, relaxing (described in Section 3 hence we will not repeat here) and clustering.

### 4.1 Definitions

Before discussing the specific steps of this algorithm, we first introduce two concepts, substitution matrix and score function. After accomplishing multiple sequence alignment on strings, in order to obtain the motif sequence, which is considered to be the center to have the closest edit distance to all sequences, we define substitution matrix as follows.

**Definition 5.** Given a group of $m$ sequences, $\alpha = \{A_1, A_2, \ldots, A_s, \ldots, A_m\}$. A substitution matrix is a group of sequences $\alpha' = \{A'_1, A'_2, \ldots, A'_s, \ldots, A'_m\}$ generated by changing $A_s$ to $A'_s$ by enlarging every $A_s$ in $\alpha$ to the same length with place holders filling in the unmatched blanks. That is to say, all sequences in the matrix have a same length.

For each sequence $A_s$ which has not been enlarged in $\alpha$, we fill its i-th character in in $A'_s$ if it matches the i-th character in the enlarged sequences in $\alpha'$. Otherwise, we fill this i-th position in $A'_s$ with a place holder. Figure 1 shows the process of transformation. The left figure shows the original $\alpha$ with sequences of various lengths, and the right one shows $\alpha'$, in which sequences are extended to the same length.

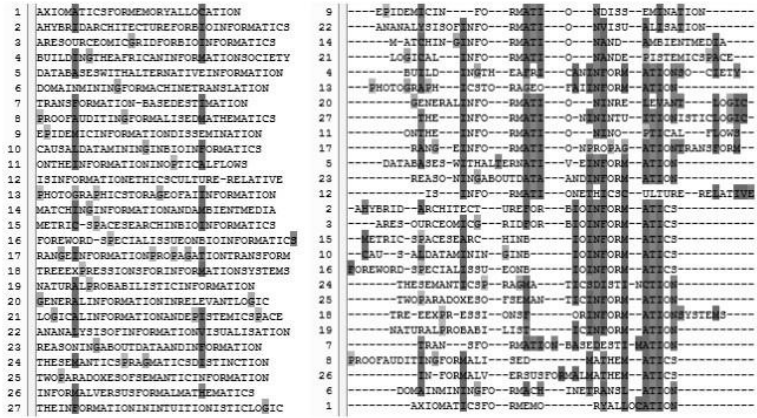

**Fig. 1.** Substitute Matrix Transformation

The method of obtaining the substitution matrix is as follows.

First, each pair of leaf nodes is compared and scored using a scoring matrix. Global optimization of dynamic programming algorithm is used in this process [30]. As for the comparison among clusters, actually, it is the comparison among groups of the multiple sequences which have already been compared. Until all sequences are processed, we obtain the substitution as a result.

To obtain the motif sequence for further selection, we need to score the sequences in the substitute matrix through a score function. This function computes scores of each kind of characters in each column of substitute matrix.

**Definition 6.** Score function, also called penalty function, is used to score the sequence alignment and generate the substitute matrix $\alpha'$ of sequences $\alpha$ and then, create motif according to $\alpha'$. The basis of score function is the scoring matrix, usually obtained by hamming distance. Higher mark represents a higher similarity among sequences.

$$dH(a, b) = \begin{cases} 0, & if\ a = b \\ 1, & else \end{cases}$$

In addition, one of the popular methods of computing scores is called *SP* (sum of pairs) standard. Take the following sequences as an example.

$$C1 = ---gttag$$

$$C2 = \text{acag} - - - \text{g}$$
$$C3 = -\text{cagttag}$$

If the bit of one sequence matches with the other one, it is marked 1, otherwise marked 0. Its mark is deducted by 1 during the inserting process. Considering the example above, we can easily find that the score of comparison between C1 and C2 is -4. Then, comparison of C1 and C3 is scored 3, and that of C2 and C3 is 0. Thus, the total score of the multiple sequences is -4+3+0=-1.

Hereby, we select the character with the highest score to fill in the corresponding position of the motif sequence. That is to say, we find characters which appear most frequently in every column to create the motif.

Generally, creating motif in this method makes a good result, except some undesirable situation when the selected sequences are far away from the motif sequence but close to each other, which negatively affects the diversity. Hence, we use $F(S, q)$ to measure the result quality. When meeting with unsatisfying results, technique mentioned in [18] is applied, where several profiles sequences or sub-sequences are used to create a more accurate motif with a cost of longer runtime. Fortunately, such special situations seldom happen in our experiment. In fact, according to triangle inequality, i.e. the sum of the length of two edges is larger than the length of the third in a triangle, two sequences cannot be too similar if they are both far away from the motif.

### 4.2 Description of Cluster

In this section, we propose the specific cluster algorithm to solve the unstable problem of genGreedy. This algorithm tries to create a motif sequence of $rlxResult$ and select results with the farthest distance to the motif for diversification.

We first treat each string as a set among which the branch length is initialized by

---

**Algorithm 3: cluster**

**Input:**
  $ED\_matrix$: two-dimensional matrix string edit distances between each pair of strings
  $rlxResult$: strings generated by Algorithm Relaxation; λ: tradeoff parameter

**Output:**
  $S$: final result set

1. set initialization ($rlxResult$)
2. **for** each set in Set **do**
3.     branchLen($set\ A, set\ B$) ← Dis($A, B$)
4. **while** ($|set|! = 1$) **do**
5.     Find $set_j$ and $set_k$ with min length of branchLen
6.         conbine $set_j, set_k$ to new $set_z$
7.         **for** each $set_t$ in Set **do**
8.             branchLen($set_z, set_t$) ← $\frac{1}{2}$ (Dis($set_t, set_j$) + Dis($set_t, set_k$))
9. $Submatrix$ ← treeSetTranstoSubstituteMatrix
10. $motif$ ← score( $Submatrix$)
11. **for** each $sequence_i$ in $Submatrix$ **do**
12.     $sequence_i.dis$ ← Dis ( $sequence_i, motif$)
13. $S$ ← $MergeSort$( $sequence_i.dis$)
14. **Return** $S$

---

$ED\_matrix$. After that, we perform search for each set in $SET$ to find two sets $set_j$ and $set_k$ that has the minimum branch length. These two sets are merged into $set_z$, and the branch lengths are updated by average distance from $set_j$ and $set_k$ to the other sets. Hence, a phylogenetic tree is built, in which closer nodes are more similar to each other. From the leaves, we start to compare the nodes. For each time, we choose two closest nodes to be added to the substitution matrix and build up this matrix by iterative processing. After that, by applying the score function, we obtain the motif sequence, which is considered to be the center with the closest edit distance to all sequences. After that, we sort the sequences by the distance to motif and return $1/(1+\lambda)*|rlxResult|$ items as the final result.

In Algorithm 3, we initialize the branch lengths among sets by edit distance among strings (Line 1 to Line 3). Line 4-Line 8 are the iterative process of building a phylogenetic tree. When two sets are merged into one, we update the branch lengths of the new set in Line 8 by computing the average of two old sets, $\frac{1}{2}$ ($\text{Dis}(set_t, set_j)$ + $\text{Dis}(set_t, set_k)$). In Line 9, we transfer our tree into a substitution matrix by multiple sequence alignment and use the score function to find the motif (Line 10). Thereafter, we calculate the distance among sequences to motif (Line 12) and return those with the farthest distance (Line 13 and 14). In this algorithm, the time complexities of 'while' loop, merge sort and other lines are $O(k^2 \log k)$, $O(k \log k)$ and $O(k)$, respectively. Thus, the total time complexity is $O(k^2 \log k)$, in which $k$ is the size of $rlxResult$.

## 5  CB2S

genGreedy and genCluster perform well in some cases. However, they still have disadvantages. genGreedy is unstable for greedy selection, and genCluster costs more time. Motivated by this, we develop a novel algorithm CB2S (Cluster-Based String Search), which is stable and meanwhile, efficient. This method combines query relaxation with diversification in one iterative process instead of two separate steps of picking candidates and filtering. Therefore, it eliminates the exceptional situations that bad candidates lead to terrible results. Meanwhile, one iterative process helps to reduce algorithm runtime. Hence, the efficiency of CB2S outperforms these algorithms and is high especially when meeting with massive datasets.

Basically, this algorithm reduces the searching space by cluster analysis in advance and then searches several clusters to retrieve results that fit our requirement. This method involves a complex pretreatment process (described in Section 5.1) of cluster analysis based on *kmeans* algorithm [26]. In Section 5.2, we discuss the details of the searching process based on *knn* algorithm [27], which is used for string classification in our approach. Given some clusters of classified strings and an unclassified string $t$, KNN tells which cluster $t$ belongs to according to a training set by uniform random sampling from classified strings. For efficiency issues of clustering and classification, strings are vectored before searching.

During the whole process, we first separate dataset into clusters, and treat the sampled data of these clusters as the training set. Given a query $q$, the training set and KNN algorithm vaguely classify $q$ to one of the clusters. This cluster is considered as the search center. The search starts from the center cluster, and the search space spreads to neighbor clusters to diversify the results. This process iterates until we obtain enough number of results.

### 5.1  Data pretreatment

The pretreatment process of CB2S aims at splitting a large dataset into small clusters and generating a complete graph by treating clusters as vertexes and distances among clusters as edges. The process is divided into the following steps. First, strings are changed to vectors. This step is necessary since using feature vectors to conduct cluster and classification work is more efficient [9]. Second, the dataset is separated into clusters by *kmeans* algorithm. In this step, similar strings are clustered in the same cluster, and strings in different clusters are less alike. After that, we generate the complete graph of clusters by calculating the distance from the center of one cluster to the others'. Strings are more similar if their clusters are near, vice versa. The details of these steps are shown below.

**Vectorization.**
In the area of machine learning, text is transferred into feature vectors to perform text mining [9]. In our work, we transfer strings into vectors and classify strings by cluster analysis on vectors. Since the representation of a string has a strong impact on the accuracy of a learning system, various techniques are proposed to fit the need of various systems [10] [11]. Word stems work well on strings. Hence, similar strings have a closer space vector. For example, the distance between the vector of "computer" and "computing" should be smaller than that between "computer" and "apple" because "computer" and "computing" are mapped in the same stem. In our paper, we use the method of vectorization presented in [9] and feature selection technique proposed in [8]. Before this process, the long strings are segmented and some end words or stop words such as "the" and "a" are removed.

**Establishment of clusters.**
We cut the whole searching space into smaller pieces and search in only small clusters to reduce the cost. Given the feature vectors of long strings obtained by vectorization, we use algorithm *K-means* to cluster them into $M$ categories. In our approach, the number of categories ($M$) is determined according to the sizes of datasets. The size of each cluster is controlled to no more than 64MB, for it is the default capacity of block storage in a distributed system, considering a future optimizing work of running our algorithm on distributed platforms. We create a distance matrix by calculating the distances among the cluster centers, i.e., the median of the set. Through the process, we obtain the complete graph of $M$ clusters.

## 5.2 The searching process

During search, we first vaguely determine which category the query input $q$ belongs to, and this category will be considered as the center set for further searching. During this process of classification, we use KNN algorithm for it does not need any evaluation parameters [27]. After that, the distance matrix among clusters is checked and sorted. A set list is returned according to the ascending order of distances between center set and the other clusters.

After determining the center set, the iterative process of retrieving enough items is shown in Figure 2. We first focus the searching range on center set, adding the string whose vector is the closest to that of query $q$. This process continues until the objective function $F(S, q)$ starts to decrease, which means that the quality of result set starts to drop. Hence, we need to switch to next category to get items with better quality. The iteration ends when the number of results satisfies user's requirement.

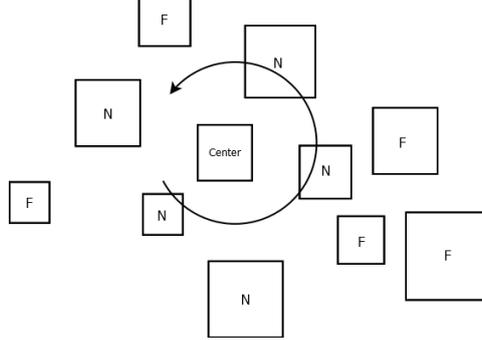

**Fig. 2.** Searching Process of CB2S

The pseudo code of the algorithm is shown in Algorithm 4. We search at least $(1 - \lambda)k_{min}$ strings for similarity and at least $\lambda k_{min}$ strings for diversity. We first use KNN algorithm to decide the center set (Line 1). In Line 2 we check the distances among sets and prune sets that are far from the center set. From Line 3-13, we search iteratively until $|S| < k_{min}$. For each round, we add the closest string to $S$ and check the change of $F(q,S)$(Line 7). When $F(q,S)$ decreases, or when $|S| > (1-\lambda)k_{min}$ is satisfied, which means that we already have enough similar strings, the searching space switches to the next cluster (Line 9). We check the number of results and jump out the loop when $|S| >= k_{max}$ to finish searching. Otherwise, we update $F(q,S)$ and continue the search processing.

We apply pruning technique when determining the search space of CB2S algorithm. The first step utilizes $k$NN algorithm to find the set of search center, to which the string of query input is vaguely classified. To reduce the search space, we abandon some of the separated sets with the lowest possibility to contain the final search result. Consider that we have $M$ sets in total, a parameter $\sigma$ is used to prune $(1-\sigma)*M$ sets that have the farthest distance to the center set. This question is converted into SSSP (single source shortest path) problem which picks $\sigma*M$ closest sets to the center set. $\sigma$ is changeable according to various datasets and the user input of $k_{min}$ and $k_{min}$. No matter how large a dataset is, the real search space includes only several clusters. For a large dataset, we set a small $\sigma$. And we set a relatively large one for small datasets to ensure enough but not redundant search space.

**Algorithm 4: CB2S**

**Input:**
>   $q$:query string; $k_{min}, k_{max}$:minimum and maximum bound; λ: tradeoff parameter
>   ε: threshold of edit distance; $Vec\_DS$: vector set of Dataset DS

**Output**:
>   $S$: final result set

1. centerSet ← KNN($q, Vec\_DS$)
2. setList ← Prune(centerSet)
3. **while**($|S| < k_{min}$) **do**
4.     **for** ($set = centerSet; set_{num} < setList.length;$) **do**
5.         add closest string $s$ to $S$
6.         delete $s$ from set
7.         $temp$ ← update $(F(q,S))$
8.         **if** $temp < F(q,S)| |S| > (1-\lambda)k_{min}$
9.             $set.$ switch
10.        **else if** $|S| >= k_{max}$
11.            **return** $S$
12.        **else**
13.            $F(q,S) \leftarrow temp$

Suppose that we have N strings in the whole dataset *DS*, and *DS* is cut into *M* small categories. The training set used to decide the center set is $N'$ which is much smaller than *N*. (In our work, we sample 5% content of each dataset randomly to do training task.) Thus The time complexity of CB2S is $O(kM + N')$, in which $k$ is the number of searching results. After applying the technique of pruning, the complexity is reduced to $O(\sigma kM + N')$, where $\sigma < 1$ and is a changeable threshold of pruning.

## 6   Experiments

In this section we evaluate the performance of our methods of optimizing the search results, and compare them with some other methods from previous papers [12].

### 6.1   Setup

We use four datasets from different domains, including Computer Conference, Information of Mammal, Protein and Random Sequence. Conference set is extracted from DBLP, containing names of journals and conferences respectively. The datasets Protein and Mammal are available on UNIPROT. Finally, the Random

Sequence is generated by ourselves, containing strings made of random combinations of letters specially used to evaluate the runtime of the algorithms. The specific statistics of all datasets are summarized in Table 1. The statistics show that the strings in Mammal and Conference are shorter than the other two datasets, usually with lengths of 50 to 100 characters. On the contrary, Protein and Random consist of relatively long strings.

**Table 1.** Dataset Characteristics

| Dataset | Number of items | Max length | Average length |
|---|---|---|---|
| Conference | 2199 | 125 | 89 |
| Protein | 10000 | 2163 | 465 |
| Mammal | 50000 | 142 | 73 |
| Random | 150194 | 572 | 277 |

All of our experiments are performed on a PC with quad-core, 64-bit, 1.7 GHz CPU and 4 GB memory. Apart from the preprocessing of CB2S which is written in Python (containing transferring strings to vectors and classification of the query input), the other parts of this system are implemented with c++. The operating system is Windows 7. Comparisons were made among our methods and two previously presented algorithms dealing with a similar problem. Performances of these five algorithms are measured through the value of $F(S, q)$ and the number of results. Efficiency is measured through runtime and the impacts of some parameters are tested by variable controlling.

### 6.2 Preprocessing Time

**Establishing index.**
Before doing similarity query, inverted table needs to be established to shorten the query time. The cost of our inverted index is $O(n^2)$. This experiment is tested on three real datasets and one synthetically generated dataset. The dataset description and the time used to build inverted table are shown in Table 2. Such is offline time, for the inverted table is only built once every time when a new dataset was read, which means that when processing other queries, we use the same inverted table.

By using such index, query processing can be very fast. We make comparison of processing query by traditional method and by using our index, the histogram Figure 4 (a) with time unit of millisecond illustrates that when doing similarity query, the runtime of using index only costs about one-third time when the dataset is not very big and the advantage can be more obvious on more complex datasets. Even though we add the query time together with time of building inverted table, the total time is still much shorter than that of the traditional method.

**Table 2.** Time of Establishing Inverted Table

| Dataset | Number of Strings | Time of establishing inverted table (s) |
|---|---|---|
| Conference | 2199 | 0.09245 |
| Protein | 10000 | 0.64021 |
| Mammal | 50000 | 0.94835 |
| Random | 150194 | 3.78483 |

**Vector transformation and dataset cluster.**
In algorithm CB2S we perform vectorization by calling the python interface 'word2vec' provided by Spark 2.0.0. with Hadoop 2.7. The development tool is py-Charm. After getting access to the vectors, we use the interface of *k-means* in MATLAB to separate datasets into clusters. The runtime of transformation in four datasets is shown below. Although the vectorization costs some time, this process does not need to be run for a second time if there is not any change in the searching space. Also, vector matrix does not need to rebuild when new items are added. Just update the vector matrix by calling the interface of increment.

Table 3. Preprocessing Time of CB2S

| Dataset | Number of Strings | Time of vectorization (s) | Time of clustering (s) |
|---|---|---|---|
| Conference | 2199 | 5.373 | 0.810 |
| Protein | 10000 | 24.694 | 1.383 |
| Mammal | 50000 | 11.025 | 1.278 |
| Random | 150194 | 41.78483 | 1.735 |

### 6.3 Impact of parameters

In this section, we tested the performance of our query processing algorithms considering two parameters that might influence the final object function $F(S,q)$, including the trade-off threshold $\lambda$, the value of average of $k_{min}$ and $k_{max}$. When making analysis of $F(S,q)$, we can clearly find the associated relationship between $F(S,q)$ and $\lambda$. However, when it comes to $k$, $k$ influences $argDiv$ and $argSim$. Hence, $k$ influences $F(S,q)$ but not directly. We fix $k$ to see how $F(S,q)$ changes with the change of $\lambda$. After that, we do the same to $k$. This part is tested in dataset Random, for this dataset is more well-distributed without too much special or extreme data. The default value of $\lambda$ is 0.5 and $k_{min}$=25, $k_{max}$=55. We set $\varepsilon$=30 and changes of $F(S,q)$ of three algorithms are shown as below.

From Figure 3, we see an increasing trend of three algorithms, which is just as what we expected. The goal function $F(S,q) = \lambda\, argDiv(S,q) + (1-\lambda)\bigl(-argSim(S,q)\bigr) = \lambda\bigl(argDiv(S,q) + argSim(S,q)\bigr) - argSim(S,q)$ when $\lambda = 0.5$, $F(S,q) = 0.5(argDiv - argSim)$. Both $argDiv$ and $argSim$ rise when $k$ gets larger. However, $argDiv$ grows faster than $argSim$, thus leads to the final trend.

The result of $F(S,q)$ changing with $\lambda$ is relatively similar. $\lambda = 0$ means traditional similarity query without considering inner diversity, while a higher value of $\lambda$ tries to involve more diversity. When $\lambda = 1$, only diversity is taken into consideration ignoring the distance to query input. Although the line of genCluster slightly decreased from $\lambda = 0.6$ to $\lambda = 0.7$, the increasing trends of three algorithms are obvious.

### 6.4 Comparisons

In this section, we compared the performance of our relaxation-diversify algorithms with two other algorithms swap and comGreedy presented in previous paper [12]. Although they are used to solve a different problem of document mining, when changing the document to be processed into dataset of strings, processing query in datasets and diversifying the query results can get a result set similar to our approaches. Thus, we choose these two algorithms for comparison.

In this experiment, we set $k_{min} = 25$ and $k_{max} = 55$ to do query with our algorithms and use $\varepsilon = 40$ as the initial edit distance threshold. The datasets used for comparison are mentioned in Sec 6.1. We use runtime to evaluate the efficiency and the value of object function $F(S,q)$, the number of results to estimate their performance.

**Efficiency**.
The runtime of five algorithms is tested on dataset Mammal, for Mammal is well-distributed and large enough. The string length covers from 40 to 200 and does not have too much special data. We set $\lambda$, $\varepsilon$ and $k$ the default values mentioned in the last section, and change the data size to see how runtime changes. The efficiency shows in Figure 4 (b) and the time unit is second.

From the figure we observe that CB2S runs fastest and comGreedy is the slowest one. CB2S puts more time on preprocessing which makes a contribution to its fast speed. Especially when running on larger datasets, the advantage of CB2S is more obvious for it effectively reduces the searching space. With the growth of data size, the runtime does not increase too much. Behind CB2S, the efficiencies of genGreedy, genCluster and swap are similar. ComGreedy is the slowest algorithm due to its complexity. The speed can also be proved by comparing the algorithm complexity of these five methods. Complexities of CB2S, genGreedy, swap, genCluster and comGreedy are $O(\sigma kM + N')$, $O(kN + k\log k)$, $O(Nk \log k)$, $O(Nk)$ and $O(Nk)$, respectively. CB2S wins in complexity. Although the comGreedy also possesses a low complexity, it takes longer time to run for multiple passes [12].

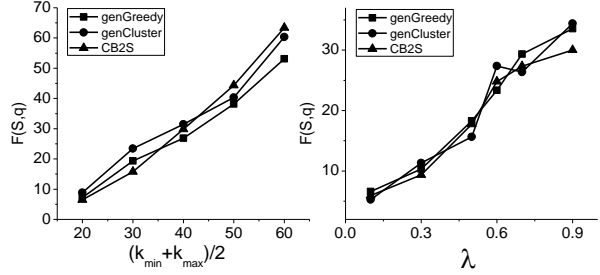

**Fig. 3.** Impacts of Parameters

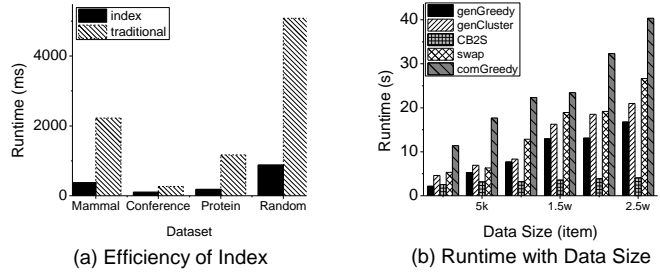

**Fig. 4.** Runtime Comparison

**Performance.**
We tested the performance through measuring the value of $F(S, q)$ and the number of results. We change the threshold of edit distance to query a specific string. Relevance of each algorithm is obtained by calculating the average of edit distance between result items and the user input, and the diversity is calculated by inner edit distance between each pair of strings. In four datasets the results appear similar in some extent.

Figure 5 illustrates the generally increasing trend of five algorithms. On different datasets, the lines fluctuate at some thresholds of $\varepsilon$. Such conditions happen when the items added to result set are not good enough but still has to be added to fit the requirement of returning $k_{min}$ to $k_{max}$ results. The gaps among algorithms are not very significant.

From Figure 6, we observe that the numbers of items in the result set generated by genCluster and genGreedy and CB2S are obviously more than the other two, nearly double in Conference, Protein and Radom. genGreedy and genCluster usually return the most results in five algorithms while CB2S is just a little fewer. These three algorithms fit the requirement of range from 25 to 55 items. We can find that comGreedy is not very stable according to variable datasets. Also, when the threshold of edit distance is small, the results returned by swap and comGreedy can be too few. In this section, our algorithms perform better.

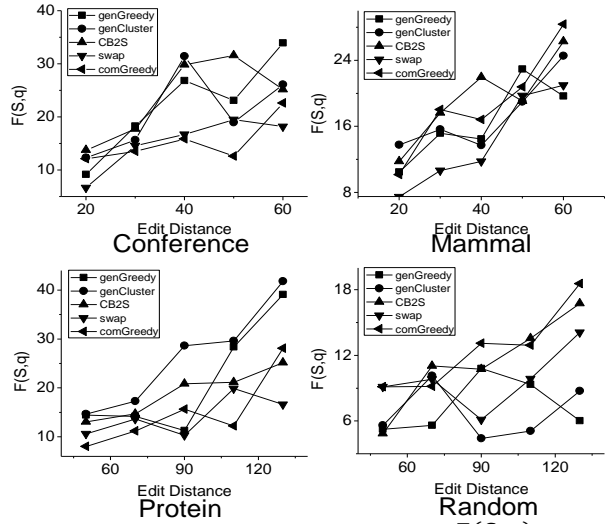

**Fig. 5.** Performance Comparison of $F(S,q)$

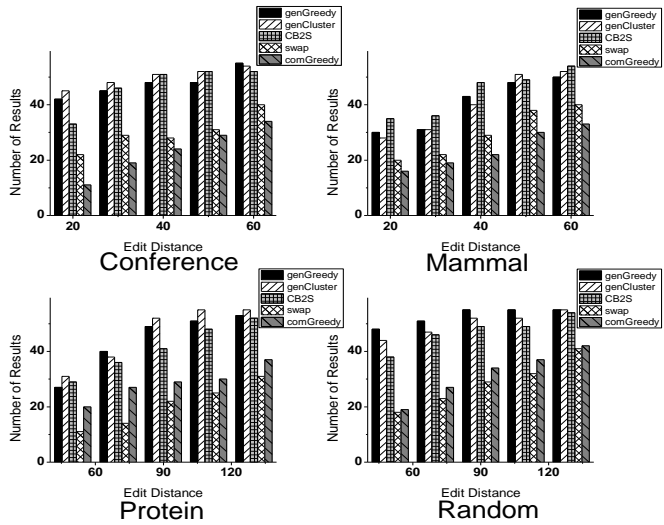

**Fig. 6.** Performance Comparison

## 7 Conclusion

To obtain high-quality results, this paper combines query relaxation and result diversification. We develop a new measure for such combination. To process query efficiently, we propose three algorithms, genGreedy, genCluster and CB2S, for different scenarios. As far as we know, this is the first work to balance query relaxation and result diversification. We evaluate our work on various datasets. The experiment shows that when providing similar relevance to query input and similar inner diversity in result set, our algorithms relax the result set to a proper size at high speed. genGreedy and genCluster are simple and effective without complex preprocessing. CB2S needs some time to do preprocessing work, but performs very well in speed, especially in large searching space. Our future work will focus on parallelizing CB2S on a distributed system to achieve higher efficiency.

***Acknowledgments.*** This paper was partially supported by NSFC grant U1509216, 61472099, National Sci-Tech Support Plan 2015BAH10F01, the Scientific Research Foundation for the Returned Overseas Chinese Scholars of Heilongjiang Province LC2016026 and MOE–Microsoft Key Laboratory of Natural Language Processing and Speech, Harbin Institute of Technology. Hongzhi Wang is the corresponding author of this paper.

# 8 References

1. C Li,J Lu,Y Lu.Efficient Merging and Filtering Algorithms for Approximate String Searches. IEEE International Conference on Data Engineering, 2008
2. Kaiping Zheng, Hongzhi Wang. A Survey of Query Result Diversification. Knowledge &Information System.2016.
3. Ziegler C N, Mcnee S M, et al. Improving recommendation lists through topic diversification. Promontory Press. 1974.
4. Drosou M, Pitoura E. DisC diversity: result diversification based on dissimilarity and coverage. Proceedings of the Vldb Endowment. 2013.
5. Agrawal R, Gollapudi S, Halverson A, et al. Diversifying search results ACM International Conference on Web Search & Data Mining. 2009.
6. Deng, D, Li, G.,Feng, J. A pivotal prefix based filtering algorithm for string similarity search. SIGMOD. 2014.
7. Jain A, Sarda P, Haritsa J R. Providing Diversity in K-Nearest Neighbor Query Results. Lecture Notes in Computer Science. 2003.
8. Yang, Yiming, Pedersen. A Comparative Study on Feature Selection in Text Categorization. Advances in Information Sciences & Service Sciences, 2012.
9. Joachims T. Text Categorization with Support Vector Machines: Learning with Many Relevant Features. Proc. European Conf. 1998.
10. Kim J D, Ohta T, Tateisi Y, et al. GENIA corpus--semantically annotated corpus for bio-text mining. Bioinformatics. 2003.
11. Larsen, BjornarAone, ChinatsuB. Fast and Effective Text Mining Using Linear-time Document Clustering. KDD-ACM.1999.
12. Yu C, Lakshmanan L, Amer-Yahia S. It takes variety to make a world: diversification in recommender systems. EDBT 2009.
13. Haveliwala T H, Gionis A, Klein D, et al. Evaluating strategies for similarity search on the web. International Conference on World Wide Web. 2010.
14. Zheng J G, Howsmon D, Zhang B, et al. Entity linking for biomedical literature. BMC Medical Informatics and Decision Making. 2015.
15. Gish W, States D J. Identification of protein coding regions by database similarity search. Nature Genetics. 1993.
16. Drosou, Marina, Pitoura, et al. Search result diversification. Proceedings of the National Academy of Sciences. 2010.
17. Vee E, Srivastava U. Efficient Computation of Diverse Query Results. 2008.
18. C. Jones and Pavel A. Pevzner. An Introduction to Bioinformatics Algorithms. MIT Press, Cambridge, 2004. Page 97-100
19. Santos L, et al. Combine-and-conquer: improving the diversity in similarity search through influence sampling ACM Symposium on Applied Computing.2015.
20. Santos, L.F.D., Oliveira, W.D., Ferreira. Parameter-free and domain-independent similarity search with diversity. SSDBM. 2013.
21. Mirzadeh N. Supporting User Query Relaxation in a Recommender System. E-Commerce and Web Technologies. 2004.
22. Zhou X, Gaugaz J. Query relaxation using malleable schemas. ACM SIGMOD, 2007.
23. Wagner R A, Lowrance R. The String-to-String Correction Problem. Journal of the Acm, 1974.
24. Zhang Z, Hadjieleftheriou M. Bed-tree: an all-purpose index structure for string similarity search based on edit distance. SIGMOD 2010.
25. Thompson J D. CLUSTAL W: improving the sensitivity of progressive multiple sequence alignment through sequence weighting, position-specific gap penalties and weight matrix choice. Nucleic Acids Research, 1994.
26. Hartigan J A, Wong M A. A K-Means Clustering Algorithm. Applied Statistics, 1979.
27. Han E H, Karypis G. Text Categorization Using Weight Adjusted k -Nearest Neighbor Classification. Pacific-Asia Conference on Knowledge Discovery and Data Mining. 2001.
28. Vargas S, Castells P. Explicit relevance models in intent-oriented information retrieval diversification.International Acm Sigir Conference on Research & Development in Information Retrieval. 2012.
29. Sun F, Wang M, Wang D, et al. Optimizing social image search with multiple criteria: Relevance, diversity, and typicality. Neurocomputing, 2012.
30. Yang J, Hu G. Computational biology: methods and applications for the analysis of biological sequences. www.sciencep.com. 2010.